\documentclass[article, eqsecnum, aps]{revtex4}

\begin{document}
\title{The Schwinger Representation of a Group: Concept and
Applications}
\author{S. Chaturvedi}
\address{ School of Physics, University of Hyderabad, Hyderabad 500 046,
India - scsp@uohyd.ernet.in}
\author{G. Marmo}
\address{Dipartimento di Scienze Fisiche, Universita di Napoli Federico II and
INFN, Via Cintia, 80126 Napoli, Italy - giuseppe.marmo@na.infn.it}
\author{N. Mukunda}
\address{Centre for High Energy Physics, Indian Institute of Science,
Bangalore~560~012, India - nmukunda@cts.iisc.ernet.in}
\author{R.Simon}
\address{The Institute of
 Mathematical Sciences, C. I. T. Campus, Chennai 600 113, India - simon@imsc.res.in}
\author{A. Zampini}
\address{Dipartimento di Scienze Fisiche, Universita di Napoli Federico II and
INFN, Via Cintia, 80126 Napoli, Italy -
alessandro.zampini@na.infn.it}
\vspace{8cm}
\begin{abstract}
 The concept of the Schwinger Representation of a finite or
 compact simple Lie group is set up as a multiplicity-free direct
 sum of all the unitary irreducible representations of the group.
 This is abstracted from the properties of the Schwinger
 oscillator construction for $SU(2)$, and its relevance in
 several quantum mechanical contexts is highlighted.  The
 Schwinger representations for $SU(2), SO(3)$ and $SU(n)$ for
 all $n$  are constructed via specific carrier spaces
 and group actions.  In the $SU(2)$ case connections to the oscillator
 construction and to Majorana's theorem on pure states for any
 spin are worked out.  The role of the Schwinger Representation in
 setting up the Wigner-Weyl isomorphism  for quantum mechanics on
 a compact simple Lie group is brought out.
\end{abstract}
\maketitle
 \section{Introduction}
 The Schwinger construction of the Lie algebra of $SU(2)$ in terms
 of the annihilation and creation operators of two independent
 quantum mechanical harmonic oscillators has been used in a wide
 variety of contexts \cite{schwinger}.  These include the physics of
 strongly correlated systems \cite{arovasaur}, quantum optics of two mode
 radiation fields \cite{ardutta}, analysis of partially coherent
 classical Gaussian Schell model beams \cite{sundar}, extension
 to all three-dimensional Lie algebras and analysis of both classical and q-deformed versions
 \cite{manko}, applications in the context of quantum computing
 \cite{ruben},
   and a new approach
 to the spin-statistics theorem \cite{berry}, to mention only a few.
 This is in addition to the elegance and relative ease with which
 many results belonging to the body of the quantum theory of
 angular momentum can be derived.

 Two important features of the Schwinger construction are economy
 and completeness.  By these we mean that the unitary
 representation $(UR)$ of $SU(2)$ that is obtained by
 exponentiating the generators contains upon reduction every
 unitary irreducible representation $(UIR)$ of $SU(2)$ exactly
 once, omitting none.  The feature of economy, i.e., simple
 reducibility, is lost when one considers the natural
 generalisation of the Schwinger construction from $SU(2)$ to
 $SU(3)$: indeed in a minimal oscillator construction that ensures
 completeness, every $SU(3)$ $UIR$ occurs with infinite
 multiplicity \cite{moshinski}. An explicit construction of a complete and multiplicity-free
 representation of $SU\left(3\right)$, via harmonic functions
 on the sphere $S^{5}$, and oscillator construction of the same representation are
 given in \cite{ruegg}.

 In the present work we abstract the two special features of the
 Schwinger $SU(2)$ construction mentioned above, and make them the
 basis of the definition of what we shall call the \emph{Schwinger
 Representation} ($SR$) for an interesting class of groups.  The  groups
 we shall mainly consider are compact Lie groups with simple Lie
 algebras, while our considerations remain meaningful for finite
 groups as well.  Both of these are of considerable importance in
 the general framework of quantum mechanics.  The precise
 definition of the $SR$ is given in the next
 Section.  Here we may stress that on account of the two
 properties of economy and completeness it may be regarded as a
 `generating representation' of the group concerned.  While these
 two features are retained, what is given up in general is any
 elementary construction in terms of oscillator operators.

 A related concept of `model representations' has been introduced
 and studied by Gelfand et al. \cite{gelfand}.  However the focus there
has been on the families of classical noncompact simple Lie
groups, and moreover on the nonunitary finite dimensional
representations of these groups.  As mentioned above, our
motivations lie in possible applications of our concept in
problems arising within the framework of quantum mechanics, where
unitarity of group representations has a special significance.

The material of this paper is arranged as follows. In Section II
we introduce the notion of the $SR$ of a group and discuss its
consequences for compact Lie groups and non compact Abelian groups
$R^n$. Further, we show that while the original Schwinger $SU(2)$
representation, and that for $SU(3)$ permit interpretation in
terms of particular induced representations, this ceases to be the
case for $SU(n)$ beyond $n=3$. In section III, we discuss the
$SU(2)$ $SR$ in a manner that anticipates generalisation later and
bring out the salient features of the carrier space thus obtained.
Section IV contains application of the construction developed in
Section III to recover the Schwinger oscillator construction for
$SU(2)$ and Majorana's representation for a spin $j$ system by
sets of points on $S^2$. In section V, we develop the $SO(3)$ $SR$
and contrast it with the way this is done conventionally. In
Section VI we show how the formalism developed in Section III for
the $SU(2)$ case naturally leads to the $SU(n)$ $SR$ for any $n$.
The significance of the $SR$ in the context of the Wigner-Weyl
isomorphism for Lie groups developed by the present authors is
brought out in Section VII. Section VIII contains concluding
remarks and some open questions which merit further investigation.

Throughout this paper we shall adopt the usual quantum mechanical
usage and denote unitary Lie group representation generators by
hermitian operators.

\section{The Schwinger Representation of a group}
We consider a compact Lie group $G$ with simple Lie algebra
$\underline{G}$.  (However many of the ideas developed below are
meaningful also for finite groups).  Then, as is well known, every
representation of $G$, and in particular every irreducible
representation, may be assumed to be unitary.  We shall use a
notation for the UIR's which generalises the notation familiar for
$SU(2)$ and $SO(3)$ in quantum angular momentum theory.  We label
the various mutually inequivalent UIR's of $G$ by a symbol or index
$j$, standing in general for a collection of independent quantum
numbers. (For $SU(2)$, $j$ is a single numerical label taking values
$0,1/2,1,3/2,\ldots$).  Within the $j^{th}$ UIR, realised on a
Hilbert space ${\mathcal {H}}^{(j)}$ of finite dimension $N_j$, we
shall write $\left(D^j_{m^{\prime}m}(g)\right)$ for the unitary
matrices representing elements $g\;\in G$ in a suitable orthonormal
basis.  The row and column indices $m^{\prime},m$ are
generalisations of the magnetic quantum number in angular momentum
theory; like $j$, they too in general stand for collections of
independent quantum numbers.  (For $SU(2), N_j=2j+1$ and $m=j,
j-1,\ldots, -j$).  In terms of a normalised translation invariant
volume element $dg$ and associated invariant delta function
$\delta(g)$ on $G$, these matrices obey the orthogonality and
completeness conditions
     \begin{eqnarray}
     \int\limits_{G}\;dg\;D^j_{mn}(g)\;D^{j^{\prime}}_{m^{\prime}
     n^{\prime}}(g)^* &=& \delta_{jj^{\prime}}\delta_{mm^{\prime}}
     \delta_{nn^{\prime}}/N_j ,\nonumber\\
     \sum\limits_{jmn}\;N_j\;D^j_{mn}(g)\;D^j_{mn}(g^{\prime})^*
     &=&\delta\left(g^{-1}g^{\prime}\right).
     \label{Dproperties}\end{eqnarray}

We now define the $SR$ of $G$ to be the simply reducible UR
     \begin{eqnarray}
     {\mathcal {D}}_0 = \sum\limits_{j}\oplus \;D^j
     \label{SRdef}\end{eqnarray}

\noindent acting on the direct sum Hilbert space
     \begin{eqnarray}
     {\mathcal {H}}_0 = \sum\limits_j\oplus\; {\mathcal {H}}^{(j)} ,
     \label{Hzero}\end{eqnarray}

\noindent the $j^{th}$ UIR $D^j$ acting on the subspace ${\mathcal
{H}}^{(j)}$ of ${\mathcal {H}}_0$.  Thus every UIR $D^j$ of $G$
occurs exactly once in this UR.  For the Lie group case, ${\mathcal
{H}}_0$ is of infinite dimension; while if $G$ is a finite group,
${\mathcal {H}}_0$ is of finite dimension.  We can set up
orthonormal bases within each ${\mathcal {H}}^{(j)}$, constituting
all together an orthonormal basis for ${\mathcal {H}}_0$, as
follows:
     \begin{eqnarray}
     {\mathcal {H}}^{(j)} &=& \mbox{Sp}\{|jm\rangle\, | j\;\mbox{fixed},\;m\;
     \mbox{varying}\},\nonumber\\
     {\mathcal {H}}_0 &=& \mbox{Sp}\{|jm\rangle |
     j\;m\;\mbox{varying}\},\nonumber\\
     \langle j^{\prime}m^{\prime}|j m\rangle &=& \delta_{j^{\prime}j}\;
     \delta_{m^{\prime}m},
     \label{Hzbasis}
     \end{eqnarray}

\noindent so that we have
     \begin{eqnarray}
     \langle j^{\prime} m^{\prime}  | {\mathcal {D}}_0(g) |j m \rangle =
     \delta_{j^{\prime}j}\;D^j_{m^{\prime}m}(g) .
     \label{Djzero}\end{eqnarray}

We give now some immediate consequences of this definition, as well
as some familiar examples.

(i)  If $G$ is abelian, each UIR is one dimensional, $N_j=1$, and
the $SR$ is the same as the regular representation acting in the
usual way (by left or by right translations which coincide) on
square integrable functions on $G$. For nonabelian $G$ the $SR$ is
always `leaner' than the regular representation since there are
always some UIR's with $N_j
> 1$.  From this point of view, the case of simple $G$ is the
exact opposite of abelian $G$: no subgroup is normal in the
former, every one is normal in the latter.  Thus for simple $G$ we
expect qualitatively that the $SR$ will be `much smaller' than the
regular representation.

(ii) When $G$ is a compact simple Lie group, we can characterize
the $SR$ in an interesting way.  In every UR of $G$, the
generators are hermitian operators obeying the commutation
relations corresponding to the Lie algebra $\underline{G}$ of $G$.
In any individual UIR, apart from the commutation relations, the
generators also obey some algebraic (symmetric polynomial)
relations characteristic of that UIR.  In ${\mathcal {D}}_0$
however no such algebraic relations are obeyed since
\underline{every} UIR is present.  In other words the generators
of the $SR$ ${\mathcal {D}}_0$ on ${\mathcal {H}}_0$ provide in a
sense a minimal faithful representation of the enveloping algebra
of $\underline{G}$: they are not subject to any algebraic
relations beyond the commutation relations.

(iii) The simple reducibility of ${\mathcal {D}}_0$  implies that
the commutant of ${\mathcal {D}}_0$ is particularly simple: any
operator $\hat{A}$ on ${\mathcal {H}}_0$ commuting with ${\mathcal
{D}}_0(g)$ for all $g$ is necessarily block diagonal, with each
entry being some numerical multiple of the unit operator:
     \begin{eqnarray}
     \hat{A}\;{\mathcal {D}}_0(g) &=&{\mathcal{D}}_0(g)\hat{A},\;
     \mbox{all}\;g\;\in\; G \Rightarrow\nonumber\\
     \hat{A} &=& \sum\limits_{j}\;_{\oplus}\; \hat{A}_j ,\nonumber\\
     \hat{A}_j &=& c_j\; 1_j ,\nonumber\\
     1_j &=& \mbox{unit operator on}\; {\mathcal{H}}^{(j)} .
     \label{Dzcomm}\end{eqnarray}

\noindent This follows from Schur's Lemma and the Wigner-Eckart
theorem. Thus this commutant is commutative.

(iv) The $SR$ concept can be extended heuristically to the
noncompact case $G=R^n$, leading to an interesting perspective
relevant to quantum mechanics.  For a quantum system with
Cartesian configuration space $Q=R^n$, corresponding to $n$
canonical Heisenberg pairs of hermitian operators $\hat{q}_r,
\hat{p}_r, r=1,2,\ldots, n$, among whom the only nonzero
commutators are
     \begin{eqnarray}
     [\hat{q}_r, \hat{p}_s] = i\;\delta_{rs},
     \end{eqnarray}

\noindent the Stone-von-Neumann theorem tells us that upto unitary
equivalence there is only one irreducible representation of these
relations.  The Hilbert space can be described via coordinate space
wave functions $\psi(\underline{q})$ or via momentum space wave
functions $\phi(\underline{p})$:
     \begin{eqnarray}
     {\mathcal{H}} = L^2(R^n) &=& \left\{\psi(\underline{q})\in
    {\mathcal{C}} \big| \parallel\psi\parallel^2 =
     \int\limits_{R^{n}} d^nq\;|\psi(\underline{q})|^2 <
     \infty \right\}\nonumber\\
     &=&\left\{\phi(\underline{p})\in{\mathcal{C}} \big|\parallel
     \phi\parallel^2 =
     \int\limits_{R^{n}}d^np\;|\phi(\underline{p})|^2 < \infty
      \right\},\nonumber\\
     \phi(\underline{p})&=& (2\pi)^{-n/2}\int\limits_{R^n}d^nq\;
     e^{-i\underline{q}\cdot\underline{p}}\;\psi(\underline{q}),
     \nonumber\\
     \parallel\phi\parallel&=& \parallel\psi\parallel ;\nonumber\\
     \left(\hat{q}_r \psi\right)(\underline{q})&=&
     q_r\;\psi(\underline{q}),\;\left(\hat{p}_r
     \psi\right)(\underline{q}) = -i\;\frac{\partial}{\partial q_r}\;
    \psi(\underline{q});\nonumber\\
    \left(\hat{q}_r\phi\right) (\underline{p})&=&
    i\;\frac{\partial}{\partial p_r}\;\phi(\underline{p}),\;
    \left(\hat{p}_r \phi\right) (\underline{p}) =
    p_r\;\phi(\underline{p}).
    \label{Schr}\end{eqnarray}

\noindent In this context, these operator actions are usually
viewed as providing us after exponentiation with the (unique)
Stone-von Neumann UIR of the $(2n+1)$ dimensional nonabelian
Heisenberg-Weyl group of phase space displacements, the generators
being $\hat{q}_r, \hat{p}_r$ and the unit operator on
${\mathcal{H}}$. However the situation can now be viewed in an
alternative manner: each real numerical $n$-dimensional momentum
vector $\underline{p}$ corresponds to a one-dimensional UIR of the
abelian group of configuration space  translations $G=R^n :
\underline{q}\rightarrow \underline{q} + \underline{a}$; as
$\underline{p}$ ranges over all of momentum space $R^n$, each such
UIR is present in ${\mathcal{H}}$ exactly once.  (Another way of
expressing this is the statement that the Cartesian momenta
$\hat{p}_r$ form a complete commuting set). Thus we can view the
kinematics of $n$-dimensional Cartesian quantum mechanics in two
ways: we have the unique Stone-von Neumann UIR  of the $(2n+1)$
dimensional nonabelian Heisenberg-Weyl group, or equally well we
have the $SR$ of the abelian group $G=R^n$ of configuration space
displacements.

(v)  The original Schwinger oscillator construction of
$\underline{SU(2)}$ leads upon exponentiation to the $SR$ of
$SU(2)$ in the sense defined above. (The $SU(2)$ notational
details will be taken up in Section III).  Each UIR of $SU(2)$ for
$j=0, 1/2, 1,\ldots$ appears exactly once.  In the case of
$SO(3)=SU(2)/{\mathcal {Z}}_2$, the distinct UIR's are usually
labelled by $\ell=0,1,2,\ldots$; these are the integer $j$ UIR's
of $SU(2)$.  The familiar UR of $SO(3)$ on square integrable
functions on $S^2$, with the simple geometric action of $SO(3)$
elements, is a realisation of the $SR$ of $SO(3)$. The reduction
into UIR's in a multiplicity-free manner is achieved, as is
familiar, by using the orthonormal basis provided by the spherical
harmonics on $S^2$.  In Sections III and IV we describe other ways
of constructing the $SR$'s of $SU(2)$ and $SO(3)$ respectively.

After these immediate properties and examples, we make some
general remarks.  Purely from the representation theory point of
view, the $SR$ ${\mathcal {D}}_0$ of $G$ is completely defined by
the statement in (\ref{SRdef},\ref{Hzero}) of its UIR content.
However, from the point of view of possible applications in the
framework of quantum mechanics, considerable interest attaches to
various ways in which this UR may be realised, with corresponding
carrier spaces and group actions.  A general way to construct UR's
of a group $G$ is by the process of induction starting from UIR's
of some subgroup \cite{mackey}. Let $H\subset G$ be some subgroup,
and $D_0$ be a UIR of $H$.  Then by an elegantly simple
construction one arrives at an induced UR ${\mathcal
{D}}_H^{(\mbox{ind},D_0)}$ of $G$: the notation indicates the
roles of $H,D_0$ and the inducing procedure. Once this UR of $G$
has been obtained, one can ask for its UIR content.  Here the main
result is the reciprocity theorem. The UR ${\mathcal
{D}}_H^{(\mbox{ind},D_0)}$ of $G$ contains the UIR $D^j$ of $G$ as
many times as $D^j$ contains $D_0$ upon restriction from $G$ to
$H$. One can now ask whether the $SR$ of $G$ arises as a
particular induced UR corresponding to some carefully chosen $H$
and $D_0$.

In the case of $SU(2)$, a natural subgroup choice is $H=U(1)$ generated
by $J_3$ in the usual notation, with eigenvalues being the magnetic quantum
number $m$.  However as a quick analysis using the reciprocity theorem shows,
we find the result:
     \begin{eqnarray}
     {\mathcal{D}}_0\;\mbox{for}\;SU(2) ={\mathcal{D}}_{U(1)}^{(\mbox{ind,0})}
      \oplus{\mathcal{D}}_{U(1)}^{(\mbox{ind,1/2})}\label{su2ind}
      \end{eqnarray}

\noindent
(Here the superscripts $0$  and $1/2$ on the right hand side indicate
the $m$ values determining the $U(1)$ UIR's used in the inducing process).
The first term on the right accounts for all the integer $j$ UIR's of
$SU(2)$, while the second term accounts for the remaining half odd integer
$j$ UIR's.  In the case of $SO(3)$ we may choose $H=SO(2)$ and then we have
     \begin{eqnarray}
    {\mathcal{D}}_0\;\mbox{for}\;SO(3)
    ={\mathcal{D}}^{(\mbox{ind},0)}_{SO(2)}\label{so3ind}
     \end{eqnarray}

\noindent So in this case the $SR$ is indeed a particular induced
representation.

For $SU(3)$ this situation continues to hold \cite{chatmuk}. Each
UIR of $SU(3)$ is labelled by a pair of independent nonnegative
integers, as $(p,q)$.  It is a fact that every UIR $(p,q)$
contains the trivial (one-dimensional) UIR of the canonical
$SU(2)$ subgroup exactly once.  Thus from the reciprocity theorem
we see that
     \begin{eqnarray}
    {\mathcal{D}}_0\;\mbox{for}\;SU(3) = {\mathcal{D}}^{(\mbox{ind},0)}_{SU(2)},
     \label{su3ind}\end{eqnarray}

\noindent
where the zero in the superscript on the right stands for the trivial $j=0$
UIR of $SU(2)$.

However this trend does not continue for $SU(n)$ beyond $n=3$
\cite{footgelfand}. In fact we show in Section VI that the $SR$ of
$SU(n)$ for $n\geq 4$ is not an induced UR corresponding to any
choice of UIR of the canonical $SU(n-1)$ subgroup of $SU(n)$.
There is thus a need to develop an alternative method to construct
the $SR$ of $SU(n)$ which works uniformly for all $n \geq 2$. This
will be done for $SU(2)$ in the next Section, for $SO(3)$ in
Section V, and for $SU(n)$ in Section VI.

\setcounter{equation}{0}
\section{The $SU(2)$ Schwinger Representation}

To set notations we begin by recalling the defining UIR and Euler
angle parametrisation of $SU(2)$ \cite{qam}.  An element $g\in
SU(2)$ is a $2 \times 2$ unitary unimodular matrix
     \begin{eqnarray*}
     g=\left(\begin{array}{cc}\xi &- \eta^*\\\eta
     &\xi^*\end{array}\right),\;\;\xi,\eta \in{\mathcal{C}} ,
     \end{eqnarray*}

     \begin{eqnarray}
     |\xi|^2 + |\eta|^2 = 1.
     \label{su2def}\end{eqnarray}

     \noindent
     The hermitian generators are $\frac{1}{2} \sigma_r$, where
     $\sigma_r$ for $r=1,2,3$ are the Pauli matrices.  The
     commutation relations are
          \begin{eqnarray}
          \left[\frac{1}{2}\sigma_r, \frac{1}{2}\sigma_s\right] =
          i\;\in_{rst}\;\frac{1}{2}\sigma_t.
          \label{pauli}\end{eqnarray}

     \noindent
     In the Euler angle parametrisation we express $g$ as a
     product of three factors:
               \begin{eqnarray}
               g(\alpha,\beta,\gamma) &=& e^{-i\alpha\sigma_3/2}\;
               e^{-i\beta\sigma_2/2}\;e^{-i\gamma\sigma_3/2}\nonumber\\
               &=&\left(\begin{array}{cc}e^{-i(\alpha+\gamma)/2}\cos
               \beta/2&-e^{-i(\alpha-\gamma)/2}\sin\beta/2\\
               e^{i(\alpha-\gamma)/2}\sin\beta/2& e^{i(\alpha+\gamma)/2}
               \cos\beta/2\end{array}\right),\nonumber\\
               \mbox{i.e.} \;\;\;\xi &=&
               e^{-i(\alpha+\gamma)/2}\cos\beta/2,\;\;\;\;\;\;
               \eta=e^{i(\alpha-\gamma)/2}\sin\beta/2.
               \label{su2euler}\end{eqnarray}

     \noindent
     The ranges for $\alpha,\beta,\gamma$ are determined by the
     condition that (except possibly on a set of measure zero)
     each element (\ref{su2def}) must occur just once.  Then one
     finds \cite{qamfoot}:
              \begin{eqnarray}
              0\leq|\xi|\leq 1\Leftrightarrow 0\leq \beta \leq \pi
              ;\nonumber\\
              0\leq \arg \xi, \;\arg \eta \leq 2\pi\;\;
              \Leftrightarrow\;\;
              0\leq \alpha\leq 2\pi,\;0\leq\gamma\leq 4 \pi.
              \label{eulerranges}\end{eqnarray}

     \noindent
     The elements $g(0,0,\gamma)$ for $0\leq\gamma\leq 4\pi$
     constitute the diagonal $U(1)$ subgroup of $SU(2)$.  Since
     $\alpha$ and $\beta$ can be interpreted as azimuthal and
     polar angles on $S^2$, the form for $g(\alpha,\beta,\gamma)$
     in (\ref{su2euler})) is in manifest agreement with the statement
     $SU(2)/U(1)=S^2$.  The normalised invariant volume element is
          \begin{eqnarray}
          dg = d\alpha\sin\beta d\beta\cdot d\gamma/16\pi^2.
          \label{su2haar}\end{eqnarray}

     The unitary representation matrices in the $j$th UIR are, as
     is familiar \cite{qamfoot1}:
          \begin{eqnarray}
          <jm|\;D^j(\alpha,\beta,\gamma)|jn> &\equiv& D^j_{mn}
          (\alpha,\beta,\gamma)\nonumber\\
          &=&e^{-im\alpha-in\gamma} d^j_{mn}(\beta)
          \label{wignersu2}\end{eqnarray}

     \noindent
     with $d^j_{mn}(\beta)$ real.  In verifying the orthogonality
     relation
          \begin{eqnarray}
          \int\limits_{SU(2)}dg\;D^j_{mn} (\alpha,\beta,\gamma)
          \;D^{j^{\prime}}_{m^{\prime}n^{\prime}}(\alpha,\beta,\gamma)^*
          =\delta_{jj^{\prime}}\delta_{mm^{\prime}}\delta_{nn^{\prime}}/(2j+1),
          \label{su2ortho}\end{eqnarray}

     \noindent
     it is necessary to keep in mind the asymmetry between
     $\alpha$ and $\gamma$ in (\ref{eulerranges}).  Thus it is simplest to
     first carry out the $\gamma$ integration producing the
     factor $\delta_{nn^{\prime}}$.  This implies that
     $j^{\prime}-j$ and $m^{\prime}-m$ are both integral.  Then
     doing the $\alpha$ integration second leads to
     $\delta_{mm^{\prime}}$; and finally the $\beta$ integration
     produces $\delta_{jj^{\prime}}$.

     The two regular representations of $SU(2)$ act on the Hilbert
     space ${\mathcal{H}}$ of square integrable functions on
     $SU(2)$ \cite{qamfoot2}:
          \begin{equation}
          {\mathcal{H}} = \left\{\psi(\alpha,\beta,\gamma)\in {\mathcal
          {C}}\big|\parallel\psi\parallel^2\right.
          \,=\,\left.\frac{1}{16\pi^2}\int\limits^{4\pi}_{0}d\gamma\;
          \int\limits^{2\pi}_{0}d\alpha\;\int\limits^{\pi}_{0}\sin\beta
          d\beta\right.\left.|\psi(\alpha,\beta,\gamma)|^2<\infty\right\}.
          \label{Hilsu2}\end{equation}

     \noindent
     When convenient we write $\psi(g)\ldots$ instead of
     $\psi(\alpha,\beta,\gamma)$.  The left regular representation
     of $SU(2)$ is given by unitary operators
     $U(g^{\prime}),g^{\prime}\in SU(2)$, acting on $\psi$ as
          \begin{eqnarray}
          (U(g^{\prime})\psi)(g) = \psi\left(g^{\prime -1}g\right).
          \label{leftr}\end{eqnarray}

     \noindent
     Similarly the right regular representation is given by
     unitary operators $\tilde{U}(g^{\prime})$:
          \begin{eqnarray}
          (\tilde{U}(g^{\prime})\psi)(g) = \psi(gg^{\prime}).
          \label{rightr}\end{eqnarray}

     \noindent
     They obey
          \begin{eqnarray}
          U(g^{\prime})U(g)&=& U(g^{\prime}g),\nonumber\\
          \tilde{U}(g^{\prime})\tilde{U}(g)&=&\tilde{U}(g^{\prime}g),\nonumber\\
          \tilde{U}(g^{\prime})U(g)&=&U(g)\tilde{U}(g^{\prime}).
          \label{leftright}\end{eqnarray}

     \noindent
     The generators $J_r$ of $U(g)$ such that
          \begin{eqnarray}
          U(g(\alpha,\beta,\gamma)) = e^{-i\alpha J_3}
          e^{-i \beta J_2} e^{-i\gamma J_3}
          \label{sugen}\end{eqnarray}

     \noindent
     are
          \begin{eqnarray}
          J_1&=&i\left(\cos\alpha\cot\beta
          \frac{\partial}{\partial \alpha}
          +\sin\alpha\frac{\partial}{\partial\beta} -
          \frac{\cos\alpha}{\sin\beta}\;\frac{\partial}{\partial\gamma}\right),
          \nonumber\\
          J_2&=&i\left(\sin\alpha\cot\beta\frac{\partial}{\partial\alpha}
          -\cos\alpha\frac{\partial}{\partial\beta} -
          \frac{\sin\alpha}{\sin\beta}\;\frac{\partial}{\partial\gamma}\right),
          \nonumber\\
          J_3&=& -i\;\frac{\partial}{\partial\alpha}.
          \label{leftgen}\end{eqnarray}

     \noindent
     Similarly the generators $\tilde{J}_r$ of $\tilde{U}(g)$ are
          \begin{eqnarray}
          \tilde{J}_1&=&i\left(\frac{-\cos\gamma}{\sin\beta}\;\frac{\partial}
          {\partial\alpha} +\sin\gamma\frac{\partial}{\partial\beta}
          +\cos\gamma\cot\beta\frac{\partial}{\partial\gamma}\right),\nonumber\\
          \tilde{J}_2&=&i\left(\frac{\sin\gamma}{\sin\beta}\;
          \frac{\partial}{\partial\alpha}
          +\cos\gamma\frac{\partial}{\partial\beta}
          -\sin\gamma\cot\beta\frac{\partial}{\partial\gamma}\right),\nonumber\\
          \tilde{J}_3 &=&i\;\frac{\partial}{\partial\gamma}.
          \label{rightgen}\end{eqnarray}

     \noindent
     The complete set of commutation relations among them is
          \begin{eqnarray}
          \protect[J_r,J_s\protect]&=& i\;\in_{rst}\;J_t,\;
          \nonumber \\ \protect[\tilde{J}_r,
          \tilde{J}_s\protect]&=&i\;\in_{rst}\;\tilde{J}_t,\nonumber\\
          \protect[J_r, \tilde{J}_s\protect]&=&0 .
          \label{comcr}\end{eqnarray}

     \noindent
     Thus the left representation generators are right translation
     invariant and vice versa.  As is well known these two sets of
     generators share a common Casimir invariant, and are related
     by the adjoint UIR of $SU(2)$, namely the defining
     representation of $SO(3)$:
          \begin{eqnarray}
          J^2 &=& J_r J_r = \tilde{J}_r \tilde{J}_r ,\nonumber\\
          \tilde{J}_r &=& -R_{sr} (\alpha,\beta,\gamma) J_s.
          \label{cas}\end{eqnarray}

     \noindent
     Acting on $D^j_{mn} (\alpha,\beta,\gamma)$ we have:
         \begin{eqnarray}
         J_3\;D^j_{mn} (\alpha,\beta,\gamma) &=& -m\;D^j_{mn}
         (\alpha,\beta,\gamma),\nonumber\\
         \tilde{J}_3\;D^j_{mn} (\alpha,\beta,\gamma)&=&n\;D^j_{mn}
         (\alpha,\beta,\gamma),\nonumber\\
         J^2\;D^j_{mn}(\alpha,\beta,\gamma) &=&
         j(j+1)D^j_{mn}(\alpha,\beta,\gamma).
         \label{su2D}\end{eqnarray}

     We now develop a method to extract the $SR$ of $SU(2)$ from the (left) regular
     representation, in a way which generalises to all $SU(n)$.
     The functions $(2j+1)^{1/2} D^{j}_{mn} (\alpha,\beta,\gamma)$
     for all $j,m,n$ form an orthonormal basis for ${\mathcal{H}}$ in
     which the two commuting UR's $U(g),\tilde{U}(g)$ are
     simultaneously reduced into UIR's.  In the UR $U(g)$ each UIR
     $j$ of $SU(2)$ occurs $(2j+1)$ times, and the quantum number
     $n$, eigenvalue of $\tilde{J}_3$, acts as a multiplicity
     index.  (Conversely, $m$ plays this role for the reduction of
     $\tilde{U}(g)$).  We can then see that if we restrict
     ourselves to the subset of basis functions
     $D^j_{mj}(\alpha,\beta,\gamma)$ with maximum possible value
     $j$  for the eigenvalue $n$ of $\tilde{J}_3$, and to the
     subspace of ${\mathcal{H}}$ spanned by these functions, we pick up
     each UIR of $SU(2)$ exactly once from the reduction of
     $U(g)$.  This leads to the identification of a subspace
     ${\mathcal{H}}_0\subset {\mathcal{H}}$ by the definition
          \begin{eqnarray}
          {\mathcal{H}}_0=\left\{\psi(\alpha,\beta,\gamma)\in {\mathcal{H}}|
          \left(\tilde{J}_1 +
          i\;\tilde{J}_2\right)\psi(\alpha,\beta,\gamma)=0\right\}
          \label{H0right}\end{eqnarray}

\noindent (Strictly speaking, wave functions in the domain of and
annihilated by $\tilde{J}_1 + i\;\tilde{J}_2$ form a dense set in
${\mathcal{H}}_0$, which upon completion gives ${\mathcal{H}}_0$).
On the other hand we know in advance that
      \begin{eqnarray}
      {\mathcal{H}}_0 &=& \mbox{Sp}\left\{(2j+1)^{1/2}
      D_{mj}^j(\alpha,\beta,\gamma),\right.\nonumber\\
      && \left.j=0,1/2, 1,\ldots,
      m=j,j-1,\ldots,-j\right\}
      \label{H0sp}\end{eqnarray}

\noindent The equivalence of (\ref{H0right}) and (\ref{H0sp}) can
be directly established as follows.

The condition defining wave functions in ${\mathcal{H}}_0$ reads
     \begin{eqnarray}
     \left(\frac{i\partial}{\partial\gamma} - \tan\beta
     \frac{\partial}{\partial\beta}
     -\frac{i}{\cos\beta}\;\frac{\partial}{\partial\alpha}\right)\;
     \psi(\alpha,\beta,\gamma)=0.
     \label{H0equ}\end{eqnarray}

\noindent This is a complex first order partial differential
equation whereas $\alpha\beta\gamma$ are all real.  Therefore we
cannot conclude that $\psi(\alpha,\beta,\gamma)$ is effectively
reduced to a function of two independent real combinations of
$\alpha\beta\gamma$. Essentially, this is like imposing the
Cauchy-Riemann equations - $\left(\frac{\partial}{\partial
x}+i\frac{\partial}{\partial y}\right)f\left(x,y\right)=0$ - on a
complex function of two real variables. The result is that
$f\left(x,y\right)$ has to be an analytic function of the complex
combination $z=x+iy$. Considering first combinations of $\alpha$
and $\beta$, and then of $\gamma$ and $\beta$, which obey
(\ref{H0equ}), we find that $\psi(\alpha,\beta,\gamma)$ can be any
analytic function of $e^{i\alpha}\tan\beta/2$ and
$e^{-i\gamma}\sin\beta$. (The analyticity condition arises because
the complex conjugate combinations
$e^{-i\alpha}\tan\beta/2,\;e^{i\gamma}\sin\beta$ do not obey
(\ref{H0equ})). However this is equivalent to the statement that
$\psi(\alpha,\beta,\gamma)$ must be an analytic function of $\xi,
\eta$ of (\ref{su2euler}):
     \begin{eqnarray}
     \psi\in{\mathcal{H}}_0\Leftrightarrow
     \psi(\alpha,\beta,\gamma)=f(\xi,\eta).
     \label{psisol}\end{eqnarray}

\noindent On the other hand the functions
$D^j_{mj}(\alpha,\beta,\gamma)$ are known to be given by
\cite{qamfoot1}:
     \begin{eqnarray}
     D^j_{mj}(\alpha,\beta,\gamma) &=&
     \sqrt{2j!} u_{jm} (\xi,\eta),\nonumber\\
     u_{jm}(\xi,\eta)&=& \xi^{j+m} \eta^{j-m}/\sqrt{(j+m)!(j-m)!},
     \label{Duwigner}\end{eqnarray}

\noindent so the equivalence of (\ref{H0right}) with (\ref{H0sp})
follows.

To cast the UIR's present in ${\mathcal{H}}_0$ into the standard
forms of quantum angular momentum theory, we notice from
(\ref{su2D}) that the eigenvalue of $J_3$ is $-m$, and as a short
calculation shows:
     \begin{eqnarray}
     (J_1+i\;J_2) D^j_{mj} (\alpha,\beta,\gamma) =
     -\sqrt{(j+m)(j-m+1)}\;D^j_{m-1,j}(\alpha,\beta,\gamma).
     \label{crealeft}\end{eqnarray}

\noindent If we therefore define the family of wave functions
     \begin{eqnarray}
     {\mathcal {Y}}_{jm}(\alpha,\beta,\gamma) &=&(-1)^{j-m}(2j+1)^{1/2}
     D^j_{-m,j} (\alpha,\beta,\gamma)\nonumber\\
     &=&\sqrt{(2j+1)!}\;\eta^{j+m}(-\xi)^{j-m}\big/\sqrt{(j+m)!(j-m)!}
     \nonumber\\
     &=&\sqrt{(2j+1)!}\;u_{jm}(\eta, -\xi),\nonumber\\
     j=0,1/2,1,\ldots,&& m=j,j-1,\ldots,-j,
     \label{scriptY}
     \end{eqnarray}

\noindent they form an orthonormal basis for ${\mathcal{H}}_0$,
     \begin{eqnarray}
     \frac{1}{16\pi^2}\;\int\limits^{\pi}_{0}\sin\beta d\beta\;
     \int\limits^{2\pi}_{0} d\alpha\;\int\limits^{4\pi}_{0}d\gamma
     \; {\mathcal{Y}}_{jm}(\alpha,\beta,\gamma)\;{\mathcal
     {Y}}_{j^{\prime}m^{\prime}}(\alpha,\beta,\gamma)^*
     =\delta_{jj^{\prime}}\delta_{mm^{\prime}};
     \label{Yortho}
     \end{eqnarray}

\noindent and moreover for each fixed $j$, the ${\mathcal
{Y}}_{jm}(\alpha,\beta,\gamma)$ transform under the left regular
representation according to the standard form of the $j$th UIR of
$SU(2)$.  The restriction of the left regular representation from
${\mathcal{H}}$ to ${\mathcal{H}}_0$ may be denoted by ${\mathcal
{D}}_0$, and it is a realisation of the $SR$ of $SU(2)$.

The following comments may be made concerning the specific way in
which the carrier space above has been obtained.  It is important
to notice that each basis function ${\mathcal
{Y}}_{jm}(\alpha,\beta,\gamma)$ retains a dependence on each of
the three real independent arguments.  This can be easily seen
when verifying the orthonormality condition (\ref{Yortho}): doing
the $\gamma$ integration first produces $\delta_{jj^{\prime}}$,
the $\alpha$ integration next produces $\delta_{mm^{\prime}}$,
while the final $\beta$ integration produces the correct
normalisation. This is similar to the comments made earlier in
connection with eqn.(\ref{su2ortho})). This means that the
extraction of the subspace ${\cal H}_0$ within the space ${\cal
H}=L^2(SU(2))$ carrying the regular representations, since it
involves limiting oneself to solutions of a \underline{complex}
differential equation, does not amount to limiting oneself to
functions defined on a lower dimensional submanifold of the full
`configuration space' $SU(2)$. In other words, the limitation to a
subspace at the vector space level is not achieved by a limitation
to any submanifold of the group manifold. This is similar to the
relationships among the position, momentum and Bargmann
representations of the Heisenberg canonical commutation relations
in quantum mechanics. While the first two can be handled in the
real realm via the concept of polarisation of a symplectic
structure, the third brings in complex quantities in a novel
manner.

Moreover, to further clarify the meaning of the functions
$\mathcal{Y}_{jm}\left(\alpha,\beta,\gamma\right)$,namely that
they essentially depend on the three variables, and that obtaining
the $SR$ from the left regular representation does not require to
quotient the group manifold, it is possible to study their
relations with the properties of the generalised coherent states
for the group $SU\left(2\right)$. As it is well known
\cite{perelomov}, if the fiducial vector in each finite
dimensional UIR of $SU\left(2\right)$ is chosen to be the highest
weight in the Cartan-Weyl setting, then the coherent states are in
correspondence with points of a 2-sphere $S^{2}\sim
SU\left(2\right)/U\left(1\right)$, where, with the standard
identification, $\gamma$ has been quotiented away:
    \begin{equation}
    \langle
    j,m\mid\alpha\beta\rangle\,=\,D^{j}_{mj}\,\left(\alpha,\beta,\gamma=0\right)
    \label{su2CS}\end{equation}

\noindent So that the functions
$\mathcal{Y}_{jm}\left(\alpha,\beta,\gamma\right)$ are, by a
direct check:
    \begin{equation}
    \mathcal{Y}_{jm}\,\left(\alpha,\beta,\gamma\right)\,=
    \,e^{-i\gamma
    j}\,\langle
    j,m\mid\alpha,\beta\rangle\,\label{scriptYCS}
    \end{equation}

\noindent This shows, once more, that $\mathcal{Y}_{jm}$ functions
do depend on the three variables, so obtaining the $SR$ from the
left regular does not require to quotient the group manifold of
$SU\left(2\right)$.

Secondly in this carrier space each basis function is a single
term expression, a monomial, rather than a sum of several distinct
terms, which is the case for a general
$D^{j}_{mn}\;(\alpha,\beta,\gamma)$ and for the usual spherical
harmonics on $S^2$.  In the next Section we exploit these features
to connect this form of the $SU(2)$ $SR$ to other known results.

\setcounter{equation}{0}
\section{Applications of $SU(2)$ Schwinger Representation}

In this Section we use the construction of the previous Section to
link up to the original Schwinger oscillator operator construction
for $SU(2)$, and to the Majorana theorem on the geometrical
representation of pure states for a spin $j$ system for any $j$.

(a)\underline{The Schwinger Oscillator  construction}

The orthonormality relation (\ref{Yortho}) for the basis functions
${\mathcal {Y}}_{jm}(\alpha,\beta,\gamma)$ of ${\mathcal{H}}_0$
can be exhibited in an alternative form suggesting interesting
generalisation. Introduce two independent complex variables
$z_1,z_2$ proportional to $\eta,-\xi$:
     \begin{eqnarray}
     z_1=\rho\eta &=&
     \rho\;e^{i(\alpha-\gamma)/2}\sin\beta/2,\nonumber\\
     z_2= -\rho\;\xi &=&
     -\rho\;e^{-i(\alpha+\gamma)/2}\cos\beta/2,\nonumber\\
     |z_1|^2 + |z_2|^2&=& \rho^2,\;\;0\leq\rho\;< \infty.
     \label{zz}\end{eqnarray}

\noindent The uniform integration measure over the two complex
planes is
     \begin{eqnarray}
     d^2z_1\;d^2z_2&\equiv&|z_1|\;|z_2|\;d|z_1|\;d|z_2|\;d\;
     \arg\;z_1\;d\;\arg\;z_2\nonumber\\
     &=&\pi^2\;dg\cdot \rho^2\;d\rho^2 ,
     \label{d2z}\end{eqnarray}

\noindent where $dg$ is given in (\ref{su2haar}).  Then
(\ref{Yortho}) takes the form
     \begin{equation}
     (2j+1)!\;\int\int\frac{d^2z_1}{\pi}\;
     \frac{d^2z_2}{\pi}\;\delta(\rho^2-1)\,u_{jm}\left(\frac{z_1}{\rho},\;\frac{z_2}{\rho}\right)\;
     u_{j^{\prime}m^{\prime}}\left(\frac{z_1}{\rho},\;
     \frac{z_2}{\rho}\right)^* =
     \delta_{jj^{\prime}}\delta_{mm^{\prime}}.
     \label{deltaz}\end{equation}

\noindent Remembering that the last two factors of the integrand are
actually $\rho$-independent, and that the result on the right hand
side really arises from the integration over $SU(2)$ with measure
$dg$, we see that we can replace $\delta(\rho^2-1)$ by any (real
positive) function $f_j(\rho^2)$ subject to
     \begin{eqnarray}
     \int\limits^{\infty}_{0} d\rho^2\cdot \rho^2\;f_j(\rho^2) = 1,
     \label{measurerho}\end{eqnarray}

\noindent and then (\ref{deltaz}) will remain valid in the form
     \begin{eqnarray}
     (2j+1)!\;\int\int\;\frac{d^2z_2}{\pi}\;\frac{d^2z_2}
     {\pi}\;f_j(\rho^2)(\rho^2)^{-2j}\;u_{jm}(z_1,z_2)\;u_{j^{\prime}
     m^{\prime}}(z_1,z_2)^*=\delta_{jj^{\prime}}\delta_{mm^{\prime}}.
     \label{uortho}\end{eqnarray}

\noindent An easy and suggestive choice consistent with
(\ref{measurerho}) is
     \begin{eqnarray}
     f_j(\rho^2) = (\rho^2)^{2j}\;e^{-\rho^2}/(2j+1)! ,
     \label{efferho}\end{eqnarray}

\noindent which leads to
     \begin{eqnarray}
     \int\int\;\frac{d^2z_1}{\pi}\;\frac{d^2z_2}{\pi}\;e^{-|z_1|^2-|z_2|^2}\;
     u_{jm}(z_1,z_2)\;u_{j^{\prime}m^{\prime}}(z_1,z_2)^* =
     \delta_{jj^{\prime}}\;\delta_{mm^{\prime}} .
     \label{uorthozz}\end{eqnarray}

\noindent This is recognised to be just the Bargmann entire function
realisation of the Schwinger oscillator operator construction for
$SU(2)$, with the familiar complete system of basis functions
     \begin{eqnarray}
     u_{jm}(z_1,z_2) = z_1^{j+m}z_2^{j-m}\big/\sqrt{(j+m)!(j-m)!}
     \label{ujm}\end{eqnarray}

\noindent forming an orthonormal basis in the Bargmann Hilbert
space \cite{bargmann}.  The oscillator operators $a_1^{\dag},
a_2^{\dag}$ correspond to multiplication by $z_1,z_2$, while the
measure in (\ref{uorthozz}) is such that $a_1$ and $a_2$ act as
$\frac{\partial}{\partial z_1},\;\frac{\partial}{\partial z_2}$
    respectively.

    It is in this way that the original Schwinger oscillator
    operator construction for $SU(2)$ can be recovered from the
    $SR$ of $SU(2)$ in the form realised in
    the previous Section.

    (b)\underline{The Majorana representation for spin $j$}

    It is very well known from the theory of the Poincar\'{e} -Bloch
    sphere that each pure state of a spin 1/2 system (two level
    quantum system) can be represented in a unique fashion by a
    point on $S^2$.  Majorana's theorem generalises this to pure
    states of a spin $j$ system for any $j$ \cite{maiorana}.  We show how
    this result can be obtained immediately and transparently from
    the work of the previous Section.

    The orthonormal basis functions for the spin $j$ UIR contained
    within the $SR$ ${\mathcal {D}}_0$ of $SU(2)$,
    given in (\ref{scriptY}), are expressible in the form
          \begin{eqnarray}
          {\mathcal {Y}}_{jm} (\alpha,\beta,\gamma) &=& (-1)^{j-m}
          (2j+1)^{1/2} D^j_{-m,j}(\alpha,\beta,\gamma)\nonumber\\
          &=&\frac{(-1)^{j-m}\sqrt{(2j+1)!}}{\sqrt{(j+m)!(j-m)!}}
          \left(e^{-i(\alpha+\gamma)/2}\cos\beta/2\right)^{j-m}
          \left(e^{i(\alpha-\gamma)/2}\sin\beta/2\right)^{j+m}\nonumber\\
          &=&\sqrt{\frac{(2j+1)!}{(j+m)!(j-m)!}}\cdot
          \xi^{2j}\cdot (-1)^{j-m}\;\zeta^{j+m},\nonumber\\
          \zeta &=&\frac{\eta}{\xi} = e^{i\alpha}\tan\beta/2 .
          \label{Yxieta}\end{eqnarray}

    \noindent
    The variable $\zeta$, which can take any value in the complex
    plane since $0\leq\alpha\leq 2\pi,\;0\leq\beta\leq\pi$, is the
    result of stereographic projection applied to the sphere
    $S^2$, with the South pole as vertex, and onto the plane
    tangent to $S^2$, at the North pole.  Thus each $\zeta$
    corresponds to a unique point on $S^2$, the North and South
    poles being mapped onto $\zeta=0$ and $\infty$ respectively.
    A general vector $\psi$ within the spin $j$ UIR in ${\mathcal
    {D}}_0$ is thus of the form
         \begin{eqnarray}
         \psi &=&
         \sum\limits^{+j}_{m=-j}\;C_m\;{\mathcal
         {Y}}_{jm}\;(\alpha,\beta,\gamma)\nonumber\\
         &=&\sqrt{(2j+1)!}\;\xi^{2j}\cdot\sum\limits^{j}_{m=-j}\;
         \frac{(-1)^{j-m}}{\sqrt{(j+m)!(j-m)!}}\,C_m\;\zeta^{j+m}
         \label{psiY}
         \end{eqnarray}

    \noindent
    As it stands, this wave function is a common standard factor
    times a polynomial of degree $\leq 2j$ in the complex variable
    $\zeta$.  In the generic case with all $C_m\neq 0$, we have a
    polynomial of degree $2j$, so $\psi$ can be uniquely factored
    into the form
         \begin{eqnarray}
         \psi=\sqrt{(2j+1)!}\cdot\xi^{2j}\cdot C_j\cdot (\zeta -
         \zeta_1)(\zeta-\zeta_2)\ldots (\zeta-\zeta_{2j}).
         \label{psifact}
         \end{eqnarray}

    \noindent
    The (unordered) set of points $\zeta_1,\zeta_2,\ldots,
    \zeta_{2j}$ (some of which may coincide) corresponds to an (unordered)
     set of points on
    $S^2$, which set determines $\psi$ uniquely and vice versa
    (upto overall normalisation of $\psi$).  This is the
    celebrated Majorana result obtained transparently from the way
    the $SR$ of $SU(2)$ was constructed in
    Section III.

    In particular the importance of each ${\mathcal
    {Y}}_{jm}(\alpha,\beta,\gamma)$ being a single term expression
    should be appreciated.

    In the generic case above with all $C_m\neq 0$, none of the
    points $\zeta_1, \zeta_2,\ldots,\zeta_{2j}$ can either vanish
    or be infinite.  In the most general case, if $m_1\geq m_2$
    are the largest and smallest $m$ values for which $C_m\neq 0$,
    ie., $C_j=C_{j-1}=\ldots = C_{m_{1}+1}=0, C_{m_1} \neq 0,\ldots,
    C_{m_{2}}\neq 0, C_{m_{2}-1} = C_{m_{2}-2} =\ldots = C_{-j} =0$,
    the wave function $\psi$ has the form

         \begin{eqnarray*}
         \psi=\sqrt{(2j+1)!}\cdot \xi^{2j}\cdot (-1)^{j-m_1}\cdot
         \left(\frac{C_{m_{1}}\zeta^{m_{1}-m_{2}}}{\sqrt{(j+m_1)!(j-m_1)!}}+
         \right.
         \end{eqnarray*}

         \begin{eqnarray*}
         \left.-\frac{C_{m_{1}-1}\zeta^{m_{1}-m_{2}-1}}
         {\sqrt{(j+m_{1}-1)!(j-m_{1}+1)!}}+\right.
         \end{eqnarray*}

         \begin{eqnarray}
         \left.\ldots + \frac{(-1)^{m_{1}-m_{2}}C_{m_{2}}}
         {\sqrt{(j+m_{2})!(j-m_{2})!}}\right)\cdot\zeta^{j+m_{2}}
         \label{psifactext}\end{eqnarray}

    \noindent
    Then in the  Majorana representation of this $\psi$ by a
    constellation of points on $S^2$, we have $j-m_1$ points at the
    South pole $(\zeta=\infty),j+m_2$ points at the North pole
    ($\zeta=0$), and the remaining $m_1-m_2$ points away from both
    poles (but with coincidences permitted).

\setcounter{equation}{0}
    \section{The $SO(3)$ Schwinger Representation}

    This case can be handled by making suitable changes in the
    $SU(2)$ treatment in Section III.  The rotation matrix
    $R(\alpha,\beta,\gamma)$ in the defining (real orthogonal) UIR
    of $SO(3)$ is
         \begin{equation}
         R(\alpha,\beta,\gamma) = \left(\begin{array}{ccc}
         \cos\alpha &-\sin\alpha&0\\\sin\alpha&\cos\alpha&0\\
         0&0&1\end{array}\right)
         \left(\begin{array}{ccc}\cos\beta&0&\sin\beta\\
         0&1&0\\-\sin\beta&0&\cos\beta\end{array}\right)\nonumber\\
         \left(\begin{array}{ccc}\cos\gamma&-\sin\gamma&0\\
         \sin\gamma&\cos\gamma&0\\0&0&1\end{array}\right)\label{rotmat}
         \end{equation}

\noindent The Euler angles now have the ranges $0\leq\alpha,
\gamma\leq
    2\pi,\;0\leq\beta\leq\pi$, so the normalised volume element is
         \begin{eqnarray}
         d\;R
         =\frac{1}{8\pi^2}\;d\alpha\;\sin\beta\;d\beta\;d\gamma .
         \label{so3haar}
         \end{eqnarray}

    \noindent
    The Hilbert space carrying the left and right regular
    representations of $SO(3)$, denoted again by ${\mathcal{H}}$, is
         \begin{equation}
         {\mathcal{H}}=\left\{\psi(\alpha,\beta,\gamma)\in{\mathcal {C}}
         \big| \parallel\psi\parallel^2 =\frac{1}{8\pi^2}\;
         \int\limits^{2\pi}_{0}d\gamma\;\int\limits^{2\pi}_{0}d\alpha\;
         \int\limits^{\pi}_{0}
         \sin\beta\;d\beta\,
         \left|\psi(\alpha,\beta,\gamma)\right|^2<\infty\right\}
         \label{Hilso3}\end{equation}

\noindent The left and right regular representations of $SO(3)$
are defined in ways analogous to (\ref{leftr},\ref{rightr}) and
need not be repeated. The expressions for their generators, $L_r$
and $\tilde{L}_r$ say, are the same as in
(\ref{leftgen},\ref{rightgen}), and the commutation relations too
are repetitions of (\ref{comcr}). The complete set of orthonormal
basis functions, realising the complete reductions of both regular
representations, are $(2\ell+1)^{1/2}
D^{\ell}_{mn}(\alpha,\beta,\gamma):\ell=0,1,2,\ldots, m$ and
$n=\ell,\ell-1,\ldots,-\ell$; and $-m,n$ are eigenvalues of
$L_3,\tilde{L}_3$ respectively.

Following the same procedure as with $SU(2)$, we can isolate a
subspace ${\mathcal{H}}_0\subset {\mathcal{H}}$ carrying a
realisation of the $SR$ ${\mathcal {D}}_0$ of $SO(3)$ by
     \begin{eqnarray}
     {\mathcal{H}}_0&=&\left\{\psi(\alpha,\beta,\gamma)\in{\mathcal{H}}|
     \left(\tilde{L}_1 + i\; \tilde{L}_2\right)
     \psi(\alpha,\beta,\gamma) = 0\right\}\nonumber\\
     &=&\mbox{Sp}\left\{(2\ell +1)^{1/2}
     D^{\ell}_{m\ell}(\alpha,\beta,\gamma),\;
     \ell=0,1,2,\ldots,\; m=\ell,\ell-1,\ldots,
     -\ell\right\}\nonumber\\
     &&
     \label{so3H0}
     \end{eqnarray}

\noindent The identification of orthonormal basis functions
transforming in the standard way under the left regular action by
$SO(3)$ is (compare (\ref{Yxieta})):
     \begin{eqnarray}
     {\mathcal {Y}}_{\ell
     m}\;(\alpha,\beta,\gamma)&=&(-1)^{\ell-m}(2\ell+1)^{1/2}
     D^{\ell}_{-m,\ell}(\alpha,\beta,\gamma)\nonumber\\
     &=&\sqrt{\frac{(2\ell+1)!}{(\ell+m)!(\ell-m)!}}\;
     \left(e^{-i(\alpha+\gamma)}\cos^2\beta/2\right)^{\ell}
     \left(-e^{i\alpha}\tan\beta/2\right)^{\ell+m}.\nonumber\\
     &&\label{so3scriptY}
     \end{eqnarray}

\noindent The single term structure of these basis functions and
the dependences on all three Euler angles should again be noted.

We have pointed out in Section II that the more familiar way of
realising the $SR$ of $SO(3)$ is via the usual kinematical action
of rotations on square integrable functions on $S^2$, namely on
functions $\psi(\alpha,\beta)$ with spherical harmonics $Y_{\ell
m}(\beta,\alpha)$ as basis functions; and that this is the induced
UR ${\mathcal {D}}^{(\mbox{ind},0)}_{SO(2)}$. While this
realisation is fully equivalent in the sense of representation
theory to the realisation given above, one sees that the actual
carrier spaces and basis functions are quite different in the two
cases.  The realisation on $L^2(S^2)$ is appropriate for
discussing the orbital angular momentum of a spinless quantum
mechanical particle; that developed in this Section is appropriate
for describing the subset of states of a rigid body in quantum
mechanics in which the third component of the angular momentum
referred to body axes always has maximal value.

It is important to note that the Schwinger oscillator operator
construction for the group $SO(3)$ can be obtained from that of
$SU(2)$ outlined in the previous section.

Restricting the basis system in (\ref{ujm}) to the set of even
functions:
   \begin{equation}
   u_{jm}\left(-z_{1},-z_{2}\right)\,=\,
   u_{jm}\left(z_{1},z_{2}\right)\label{evenujm}
   \end{equation}
is equivalent to allow only for integer values of $j$, so to
define a space supporting a realisation of $SO(3)$ Lie algebra in
terms of oscillators. This means that the Schwinger oscillator
construction for $SU(2)$ goes through for $SO(3)$.

\setcounter{equation}{0}
\section{The Schwinger Representation for $SU(n)$}

We now show how the $SU(2)$ procedure developed in Section III can
be extended to the entire family of unitary unimodular groups
$SU(n)$.  We begin with preliminaries about $\underline{SU(n)}$,
then prove that for $n\geq 4$ the $SR$ of $SU(n)$ cannot be
obtained by the inducing construction from any UIR of the
canonical $SU(n-1)$ subgroup.  We then sketch the generalisation
of the $SU(2)$ procedure to general $SU(n)$, and give details in
the $SU(3)$ case.

In the so-called tensor notation the Lie algebra
$\underline{SU(n)}$ of $SU(n)$ consists of operators
$A^{\lambda}\;_{\mu},\;\;\lambda,\mu=1,2,\ldots,n$, obeying the
commutation, conjugation and algebraic relations \cite{okubo}:
     \begin{eqnarray}
     [A^{\lambda}\;_{\mu},\;A^{\rho}\;_{\sigma}]&=&
     \delta^{\rho}_{\mu}\;A^{\lambda}\;_{\sigma} - \delta^{\lambda}
     _{\sigma}\;A^{\rho}\;_{\mu},\nonumber\\
     \left(A^{\lambda}\;_{\mu}\right)^{\dag}
     &=&A^{\mu}\;_{\lambda},\nonumber\\
     A^{\lambda}\;_{\lambda}&=& 0 .
     \label{suncr}\end{eqnarray}

\noindent The subset of commuting hermitian generators which can
be assumed to be simultaneously diagonal in any UR of $SU(n)$ may
be taken to be (upto overall multiplicative factors):
     \begin{eqnarray*}
     A^1\;_1 - A^2\;_2, \;A^1\;_1 + A^2\;_2 -
     2A^3\;_3,\ldots,
     \end{eqnarray*}
     \begin{eqnarray}
     A^1\;_1 +A^2\;_2 +\ldots +A^{n-1}\;_{n-1} -(n-1)A^n\;_n =
     -n\;A^n\;_n.
     \label{suncartan}
     \end{eqnarray}

Since $SU(n)$ has rank $(n-1)$, there are $(n-1)$ fundamental
UIR's; a general UIR is obtained by forming the direct product of
several copies of each fundamental UIR and then isolating the
`largest' irreducible piece.  The fundamental UIR's are the
defining $n$-dimensional UIR consisting of $n\times n$ unitary
unimodular matrices, followed by antisymmetric tensor
representations of successive ranks $2,3,\ldots, (n-1)$ over the
defining UIR.  For brevity denote the fundamental UIR of $SU(n)$
given by antisymmetric tensors of rank $p$ by
$\underline{p}^{(n)}$, for $p=1,2,\ldots, n-1$.  Under complex
conjugation we have
     \begin{eqnarray}
     \underline{p}^{(n)^{*}} = \underline{(n-p)^{(n)}}.
     \label{sunconj}
     \end{eqnarray}

\noindent Then the reduction of each fundamental UIR under the
canonical $SU(n-1)$ subgroup is easily seen to have the two-term
structure
     \begin{eqnarray}
     \underline{p}^{(n)} = \underline{p}^{(n-1)} \oplus
     \underline{(p-1)}^{(n-1)},\; p=1,2,\ldots,n-1.
     \label{sun2term}
     \end{eqnarray}

\noindent One sees from this that for $n\geq 4$, there is no single
UIR of $SU(n-1)$ which occurs exactly once in each fundamental UIR
of $SU(n)$, hence also none which appears exactly once in each UIR
of $SU(n)$. For example, when $n=4$, we have in terms of
dimensionalities $\underline{1}^{(4)}=4,
\underline{2}^{(4)}=6,\;\underline{3}^{(4)}= 4^*;\;\mbox{their}\;
SU(3)\;\mbox{contents are}$
     \begin{eqnarray}
     4 &=& 3 \oplus 1,\nonumber\\
     6 &=& 3^*\oplus 3,\nonumber\\
     4^* &=& 1\oplus 3^* ,
     \label{su3sum}
     \end{eqnarray}

\noindent where $\underline{1}^{(3)} = 3,\;\underline{2}^{(3)} =
3^*$; and the statement made above is seen to be true.  For the
$SU(3)\rightarrow SU(2)$ case we have in contrast
     \begin{eqnarray}
     3 &=&2\oplus 1,\nonumber\\
     3^* &=& 2\oplus 1,
     \label{su3su2}\end{eqnarray}

\noindent and in fact, as mentioned in Section II, each UIR of
$SU(3)$ does contain exactly one $SU(2)$ invariant state.  From
the reciprocity theorem we conclude that for $n\geq 4$, the $SR$
of $SU(n)$ cannot be obtained by the inducing construction
starting from any UIR of $SU(n-1)$.

The method used for $SU(2)$ in Section III, however, does work for
all $SU(n)$.  The Hilbert space carrying the two commuting regular
representations of $SU(n)$ is ${\mathcal{H}}=L^2(SU(n))$:
     \begin{eqnarray}
     {\mathcal{H}}=\{\psi(g)\in{\mathcal {C}}|g\in
     SU(n),\parallel\psi\parallel^2 = \int dg|\psi(g)|^2 <
     \infty\}.
     \label{Hilsun}
     \end{eqnarray}

\noindent Here $dg$ is the normalised invariant volume element on
$SU(n)$, and the left and right regular representation operators
$U(g),\tilde{U}(g)$ are defined exactly as in
(\ref{leftr},\ref{rightr}). Let us denote their generators by
$A^{\lambda}\;_{\mu},\;\tilde{A}^{\lambda}\;_{\mu}$: each set
obeys eqns.(6.1), and they mutually commute.  Then the subspace
${\mathcal{H}}_0$ supporting a $SR$ ${\mathcal {D}}_0$ of $SU(n)$
is identified by
     \begin{eqnarray}
     {\mathcal{H}}_0&=&\left\{\psi(g)\in{\mathcal
     {H}}|\tilde{A}^{\lambda}\;_{\mu}\;\psi=0,\;\lambda <
     \mu\right\}\nonumber\\
     &=&\left\{\psi(g)\in{\mathcal
     {H}}|\tilde{A}^{\lambda}\;_{\lambda+1}\;\psi=0,\;
     \lambda=1,2,\ldots,n-1\right\}.
     \label{H0n}
     \end{eqnarray}

\noindent Here we use the fact that the $\frac{1}{2}n(n-1)$
nonhermitian operators $\tilde{A}^{\lambda}\;_{\mu}$ for $\lambda
< \mu$ close under commutation, so we can consistently look for
their common null space.  (In the defining UIR of $SU(n)$, these
are lower triangular matrices).  Since
$\left[\tilde{A}^{\lambda}\;_{\lambda+1},
\tilde{A}^{\lambda+1}\;_{\lambda+2}\right] =
\tilde{A}^{\lambda}\;_{\lambda+2}$ etc., we can adopt the more
economical definition in the second line of (\ref{H0n}).  These
conditions have the following effect: out of the many appearances
of each $SU(n)$ UIR in the reduction of the left regular
representation $U(g)$ on ${\mathcal{H}}$, exactly one is picked up
corresponding to the highest weight with respect to the right
regular representation $\tilde{U}(g)$.  Then the UR $U(g)$ on
${\mathcal{H}}$, when restricted to ${\mathcal{H}}_0$, gives a
realisation of the $SR$ ${\mathcal {D}}_0$ of $SU(n)$.

We spell out the details in the $SU(3)$ case \cite{UIRSU3}. The
$SU(2)$ subgroup is taken to be generated by $A^1\;_2, A^2\;_1,
A^1\;_1 -A^2\;_2$. In the standard isospin notation we have:
     \begin{eqnarray}
     I_3 = A^1\;_1 - A^2\;_2,I_+ = \sqrt{2}\;A^1\;_2,\;I_{-} =
     \sqrt{2}\;A^2\;_1 .
     \label{su3cartan}
     \end{eqnarray}

\noindent A general $SU(3)$ UIR is denoted by $(p,q)$, with $p$ and
$q$ independent nonnegative integers.  ($(1,0)=3=\mbox{defining
representation}, (0,1)=3^*$).  Within this UIR, whose dimension is
$N_{p,q}=\frac{1}{2}(p+1)(q+1)(p+q+2)$, an orthonormal basis is
written as
     \begin{eqnarray}
     |p,q;\;I,\;I_3,\;Y\rangle ,
     \label{su3basis}
     \end{eqnarray}

\noindent where $I,I_3$ are the usual $SU(2)$ UIR quantum numbers,
and the hypercharge $Y$ is the eigenvalue of $-A^3\;_3$. The
`$I-Y$ multiplets' contained in the UIR $(p,q)$ are given by the
rules:
     \begin{eqnarray}
     I&=&\frac{1}{2}(r+s),\;I_3=I,\;I-1,\ldots,-I,\nonumber\\
     Y&=&r-s +\frac{2}{3}(q-p),\nonumber\\
     r=0,1,2,\ldots,p,&&s=0,1,2,\ldots,q.
     \label{IYsu3}\end{eqnarray}

\noindent (Thus by taking $r=s=0$ we see that an $SU(2)$ singlet
state with $I=I_3=0$ is always present once).  The nonhermitian
generators $A^1\;_2, A^1\;_3, A^2\;_3$ cause the following changes
in the `magnetic quantum numbers' $I,I_3,Y$ of the basis states
(\ref{su3basis}):
     \begin{eqnarray}
     A^1\;_2:I,\;I_3,\;Y&\rightarrow&I,\;I_3+1, Y,\nonumber\\
     A^1\;_3:I,\;I_3,\;Y&\rightarrow&I\pm 1/2,\;I_3+1/2,\;Y+1,\nonumber\\
     A^2\;_3: I,\;I_3,\;Y&\rightarrow& I\pm 1/2,\;I_3-1/2,\;Y+1.
     \label{su3rais}
     \end{eqnarray}

\noindent Thus either $Y$ is increased by unity, or $Y$ is unchanged
but $I_3$ is increased by unity.  The unique basis state within
$(p,q)$ annihilated by $A^1\;_2$ and $A^2\;_3$ (hence also by
$A^1\;_3$) is then seen to be for $r=p,s=0$:
     \begin{eqnarray}
     |p,q;\;\frac{1}{2}p,\;\frac{1}{2}p,\;\frac{1}{3}(p+2q)\rangle
     \label{su3hw}
     \end{eqnarray}

\noindent With appropriate conventions this is the highest weight
state in the UIR: it has the highest possible hypercharge
value,and for this hypercharge it has the highest possible
eigenvalue for $I_3$.

Now we use this information about UIR's of $SU(3)$ to analyse the
regular representations.  These UR's are realised on $L^2(SU(3))$,
and an orthonormal basis is given in an obvious notation by the
collection of all unitary representation matrices:
     \begin{eqnarray}
     \sqrt{N_{p,q}}\;D^{(p,q)}_{I I_3Y;\tilde{I}
     \tilde{I}_3\tilde{Y}}(g) .
     \label{wignersu3}\end{eqnarray}

\noindent The subspace ${\mathcal{H}}_0$ identified in (\ref{H0n})
is thus seen to be spanned by those basis functions for which
$\tilde{I}=\tilde{I}_3=\frac{1}{2}p,\;\tilde{Y}=\frac{1}{3}(p+2q)$:
     \begin{eqnarray}
     {\mathcal{H}}_0&=&\mbox{null space of}\;\tilde{A}^1\;_2, \tilde{A}^2\;_3
     \left(\mbox{and}\;\tilde{A}^1\;_3\right)\nonumber\\
     &=&\mbox{Sp}\left\{\sqrt{N_{p,q}}\;D^{(p,q)}_{I I_3
     Y;\;\frac{1}{2}p,\;\frac{1}{2}p,
     \frac{1}{3}(p+2q)}(g)\right\},
     \label{H0su3}
     \end{eqnarray}

\noindent and we see explicitly that with respect to the left
action each UIR of $SU(3)$ occurs exactly once.  Thus the $SR$
${\mathcal {D}}_0$ of $SU(3)$ is realised on ${\mathcal{H}}_0$.

To exhibit a basis ${\mathcal {Y}}_{p,q; I I_3 Y}(g)$ for
${\mathcal{H}}_0$ which is orthonormal and transforms in the
standard `Biedenharn' manner under $SU(3)$ action, \cite{UIRSU3}
equations analogous to (\ref{scriptY}) have to be set up, but we
omit the details.

\setcounter{equation}{0}
\section{Application to the Wigner-Weyl isomorphism}

The Wigner-Weyl isomorphism (WWI) is a method to express states
and operators in the traditional Hilbert space formulation of
quantum mechanics in a classical phase space language \cite{weyl}.
Thus density matrices and general dynamical variables are
represented by corresponding $c$-number functions on phase space,
their Weyl symbols, while quantum mechanical expectation values
are calculated as integrals of products of Weyl symbols over phase
space in the manner of classical statistical mechanics.  The WWI
has been studied most extensively in the case of Cartesian quantum
mechanics when, as mentioned in Section II, the configuration
space is $Q=R^n$ and phase space is $R^{2n}$.

It has been shown elsewhere that if we consider the configuration
space to be a (compact simple) Lie group $G$, the kinematic
structure of quantum mechanics shows striking new features absent
in the Cartesian case, so the  WWI also exhibits unexpected
features \cite{mio}.  Interestingly the $SR$ of $G$ plays a role
in this context, and this will be outlined here.

The Hilbert space of wave functions is in an obvious notation
     \begin{eqnarray}
     {\mathcal{H}} = \left\{\psi(g)\in{\mathcal {C}}|g\in
     G,\parallel\psi\parallel^2 = \int\limits_{G} dg\;|\psi(g)|^2
     <\infty\right\}.
     \label{Hil}\end{eqnarray}

\noindent The left and right regular UR's act as in (\ref{leftr},
\ref{rightr}, \ref{leftright}) reinterpreted as referring to $G$.
A density operator $\hat{\rho}$ and a general dynamical variable
$\hat{A}$ are represented by their integral kernels
     \begin{eqnarray}
     \hat{\rho}\rightarrow \langle
     g^{\prime}|\hat{\rho}|g\rangle,\;
     \hat{A}\rightarrow \langle g^{\prime}|\hat{A}|g\rangle .
     \label{kernel}\end{eqnarray}

\noindent where the ideal kets $|g>$ for $g\in G$ are introduced
such that
     \begin{eqnarray}
     \psi(g)&=& <g|\psi> ,\nonumber\\
     <g^{\prime}|g>&=& \delta(g^{-1}g^{\prime}),\nonumber\\
     \int dg\;|g><g|&=& 1\;\mbox{on}\; {\mathcal{H}}.
     \label{kets}
     \end{eqnarray}

\noindent This allows us to express the actions of $U(g),
\tilde{U}(g)$ in the succinct forms
     \begin{eqnarray}
     U(g)|g^{\prime}> =
     |gg^{\prime}>,\;\tilde{U}(g)|g^{\prime}>=|g^{\prime}g^{-1}>.
     \label{lrket}\end{eqnarray}

\noindent The trace orthonormality of these unitary operators is
then immediate:
     \begin{eqnarray}
     \mbox{Tr}(U(g^{\prime})U(g)) =
     \mbox{Tr}(\tilde{U}(g^{\prime})\tilde{U}(g))=\delta(g^{\prime}g).
     \label{traceortho}
     \end{eqnarray}

The complementary `momentum' basis for ${\mathcal{H}}$ in which both
regular representations are simultaneously completely reduced into
UIR's is determined by the $D$-functions as
     \begin{eqnarray}
     |jmn> = N^{1/2}_{j}\;\int dg\;D^j_{mn}(g)\;|g>
     \label{momket}
     \end{eqnarray}

\noindent with the basic properties
     \begin{eqnarray}
     <j^{\prime}m^{\prime}n^{\prime}|jmn> &=&
     \delta_{jj^{\prime}}\;\delta_{mm^{\prime}}\;
     \delta_{nn^{\prime}},\nonumber\\
     U(g)\;|jmn>&=& \sum\limits_{m^{\prime}}\;D^j_{mm^{\prime}}\;
     (g^{-1})|jm^{\prime}n>,\nonumber\\
     \tilde{U}(g)\;|jmn>&=&\sum\limits_{n^{\prime}}\;D^j_{n^{\prime}n}
     \;(g)\;|jmn^{\prime}>.
     \label{mketprop}
     \end{eqnarray}

\noindent In the reduction of either regular representation each
UIR $j$ of $G$ occurs $N_j$ times.  In this basis $\hat{\rho}$ and
$\hat{A}$ are represented by `matrices'
     \begin{eqnarray}
     \hat{\rho}\rightarrow\langle j^{\prime}m^{\prime}n^{\prime}|
     \hat{\rho}|jmn\rangle,\;\hat{A}\rightarrow\langle
     j^{\prime}m^{\prime}n^{\prime}|\hat{A}|jmn\rangle.
     \label{wigsym}\end{eqnarray}

In this scheme the WWI can be set up  in  two equally good ways. We
describe both at this point even though only the second one will be
used later.

\noindent
\underline{Option I}

With an operator $\hat{A}$ described by kernel (\ref{kernel}) or
matrix (\ref{wigsym}) we associate the Weyl symbol
     \begin{eqnarray}
     W_{\hat{A}}
     (g;jmm^{\prime})&=&\int\int dg^{\prime}dg^{\prime\prime}\langle
     g^{\prime\prime}|\hat{A}|g^{\prime}\rangle\;D^j_{mm^{\prime}}
     (g^{\prime}g^{\prime\prime-1})
     \delta(g^{-1}s(g^{\prime},g^{\prime\prime}))\nonumber\\
     &=&\int\int
     dg^{\prime}dg^{\prime\prime}\langle g^{\prime\prime}|
     \tilde{U}(g)\hat{A}\tilde{U}(g)^{-1}|g^{\prime}\rangle\;
     D^j_{mm^{\prime}}(g^{\prime}g^{\prime\prime-1})
     \delta(s(g^{\prime},g^{\prime\prime})).\nonumber\\
     &&
     \label{weylsymbol}\end{eqnarray}

\noindent This symbol depends on a group element $g$ (coordinate
variable) and on the discrete UIR labels $jmm^{\prime}$ (momentum
variable).  It involves the function
$s(g^{\prime},g^{\prime\prime})\in G$ dependent on two arguments,
having the properties
     \begin{eqnarray}
     s(g^{\prime},g^{\prime\prime})&=&
     s(g^{\prime\prime},g^{\prime}),\nonumber\\
     s(g^{\prime},g^{\prime}) &=& g^{\prime},\nonumber\\
     s(g_1 g^{\prime}g_2,\;g_1 g^{\prime\prime} g_2)&=&
     g_1\;s(g^{\prime},g^{\prime\prime})g_2.
     \label{sconditions}
     \end{eqnarray}

\noindent A possible choice for $s(g^{\prime},g^{\prime\prime})$
is the `midpoint' of the geodesic in $G$ from $g^{\prime}$ to
$g^{\prime\prime}$.  Using (\ref{sconditions}) this solution can
be written as
     \begin{eqnarray}
     s(g^{\prime},g^{\prime\prime})=g^{\prime}s_0(g^{\prime
     -1}g^{\prime\prime}),
     \end{eqnarray}

\noindent where $s_0(g)$ is the `midpoint' of the one-parameter
subgroup connecting the identity $e\in G$ to $g$.

With this option we have under conjugation of $\hat{A}$ by
$\tilde{U},U$:
     \begin{eqnarray}
     \hat{A}^{\prime}
     =\tilde{U}(g_1)\hat{A}\tilde{U}(g_1)^{-1}&\Rightarrow&\nonumber\\
     W_{\hat{A}^{\prime}}(g;jmm^{\prime}) &=&
     W_{\hat{A}}(gg_1;jmm^{\prime});\nonumber\\
     \hat{A}^{\prime\prime}=U(g_2)^{-1}
     \hat{A}\;U(g_2)&\Rightarrow&\nonumber\\
     W_{\hat{A}^{\prime\prime}}(g;jmm^{\prime})&=&\sum\limits_{m_{1},m^{\prime}_{1}}
     \;D^j_{mm_{1}}\left(g^{-1}_{2}\right)\;
     W_{\hat{A}}\left(g_2 g;j m_1 m^{\prime}_1\right)\;
     D^{j}_{m^{\prime}_{1} m^{\prime}}(g_2).\nonumber\\
     &&\label{lrsuop}
     \end{eqnarray}

\noindent Finally for two operators $\hat{A},\hat{B}$ on
${\mathcal{H}}$ we find:
     \begin{eqnarray}
     \mbox{Tr}\;(\hat{A}\hat{B}) =\int
     dg\;\sum\limits_{jmm^{\prime}}\;N_j\;W_{\hat{A}}(g;jmm^{\prime})
     W_{\hat{B}}(g;jm^{\prime}m).
     \label{traceop}
     \end{eqnarray}

\noindent \underline{Option II}

To save on symbols we use the same notations as in Option I; in any
case we later make use only of Option II.  With $\hat{A}$ we now
associate the Weyl symbol
     \begin{eqnarray}
     W_{\hat{A}}(g;jnn^{\prime})&=&\int\int
     dg^{\prime}dg^{\prime\prime}\langle g^{\prime\prime}|\hat{A}|
     g^{\prime}\rangle\; D^j_{n^{\prime}n}
     (g^{\prime\prime-1}g^{\prime})
     \delta(g^{-1}s(g^{\prime},g^{\prime\prime}))\nonumber\\
     &=&\int\int dg^{\prime}dg^{\prime\prime}\langle g^{\prime\prime}
     |U(g)^{-1}\hat{A}\;U(g)|g^{\prime}\rangle\;D^j_{n^{\prime}n}
     (g^{\prime\prime-1}
     g^{\prime})\;\delta(s(g^{\prime},g^{\prime\prime})).\nonumber\\
     &&
     \label{weylsymbolII}
     \end{eqnarray}

\noindent Under conjugation of $\hat{A}$ we now have:
     \begin{eqnarray}
     \hat{A}^{\prime}=\tilde{U}(g_1)\hat{A}\tilde{U}(g_1)^{-1}&\Rightarrow&
     \nonumber\\
     W_{\hat{A}^{\prime}}(g;jnn^{\prime})&=&\sum\limits_{n_{1},
     n^{\prime}_{1}}\;D^j_{n_1
     n}\left(g_1^{-1}\right)\;W_{\hat{A}}\left(gg_1;jn_1n^{\prime}_1\right)\;
     D^j_{n^{\prime}n^{\prime}_1} (g_1);\nonumber\\
     \hat{A}^{\prime\prime} =
     U(g_2)^{-1}\;\hat{A}\;U(g_2)&\Rightarrow&\nonumber\\
     W_{\hat{A}^{\prime\prime}}(g;jnn^{\prime}) &=&W_{\hat{A}}
     \left(g_2g;jnn^{\prime}\right).
     \label{lrsuopII}
     \end{eqnarray}

\noindent
For the trace over ${\mathcal{H}}$:
     \begin{eqnarray}
     \mbox{Tr}(\hat{A}\hat{B}) = \int
     dg\;\sum\limits_{jnn^{\prime}}\;N_j\;W_{\hat{A}}(g;jnn^{\prime})\;
     W_{\hat{B}} (g;jn^{\prime}n).
     \label{traceopII}
     \end{eqnarray}

\noindent We stress that
(\ref{weylsymbol},\ref{lrsuop},\ref{traceop}) hold with Option I,
while (\ref{weylsymbolII},\ref{lrsuopII},\ref{traceopII}) with
Option II. The major differences are in the behaviours under
conjugation of $\hat{A}$.

Let us hereafter choose to work with Option II.  The structure of
the `momentum variables' in $W_{\hat{A}}(g;jnn^{\prime})$ suggests
that we bring in the $SR$ ${\mathcal {D}}_0(g)$ of $G$ acting on
${\mathcal{H}}_0$, as set up in
(\ref{SRdef},\ref{Hzero},\ref{Hzbasis},\ref{Djzero}). We can then
represent the Weyl symbol of $\hat{A}$ more compactly as
simultaneously a function of $g$ and a block diagonal operator on
${\mathcal{H}}_0$:
     \begin{eqnarray*}
     \hat{A}\rightarrow
     W_{\hat{A}}(g;jnn^{\prime})\rightarrow\tilde{A}(g)=
     \sum\limits_{j}\oplus\;\tilde{A}_j(g),\nonumber\\
     \end{eqnarray*}

     \begin{eqnarray}
     \tilde{A}_j(g)
     =\sum\limits_{n,n^{\prime}}\;W_{\hat{A}}(g;jnn^{\prime})\;
     |jn^{\prime})(jn|.
     \label{Ajtilde}
     \end{eqnarray}

\noindent Each $\tilde{A}_j(g)$ acts on the subspace
${\mathcal{H}}^{(j)}\subset{\mathcal{H}}_{0}$, and $\tilde{A}(g)$
acts in a block diagonal manner on ${\mathcal{H}}_0$.  For two
operators $\hat{A}$ and $\hat{B}$, traces within
${\mathcal{H}}^{(j)}$ give
     \begin{eqnarray}
     \mbox{tr}\left(\tilde{A}_j(g)\tilde{B}_j(g)\right) =
     \sum\limits_{n,n^{\prime}}\;W_{\hat{A}}
     (g;jnn^{\prime})\;W_{\hat{B}}(g;jn^{\prime}n),
     \label{tildetrace}
     \end{eqnarray}

\noindent so the general trace formula (7.16) has the form
     \begin{eqnarray}
     \mbox{Tr}(\hat{A}\hat{B}) =
     \int dg\;\sum_{j}\;N_{j}\;\mbox{tr}
     \left(\tilde{A}_j(g)\;\tilde{B}_j(g)\right).
     \label{tracetilde}
     \end{eqnarray}

\noindent It is important to recognise that the trace operation on
the right hand side is not over ${\mathcal{H}}_0$, because of the
presence of the dimensionality factors $N_j$.  We come back to this
point later.

We can now ask for the conditions on $\hat{A}$ which make its Weyl
symbol $W_{\hat{A}}(g;jnn^{\prime})$ independent of `coordinate'
$g$ and dependent only on `momenta' $jnn^{\prime}$ \cite{miofoot}.
From (\ref{weylsymbolII}) we see that $\hat{A}$ must belong to the
commutant of the operators $U(g)$ of the left regular
representation.  This means that it should be built up exclusively
from the operators $\tilde{U}(g)$ of the right regular
representation.  After elementary calculations we can state this
as a series of two-way implications:
     \begin{eqnarray}
     W_{\hat{A}}(g;jnn^{\prime})&=&\mbox{independent of}\;
     g\Leftrightarrow \nonumber\\
     U(g)\;\hat{A} &=&\hat{A}\;U(g),\;\mbox{all}\;g
     \Leftrightarrow\nonumber\\
     \langle g^{\prime\prime}|\hat{A}|g^{\prime}\rangle
     &=&
     f(g^{\prime\prime-1}g^{\prime}),\;\mbox{some}\;
     f\Leftrightarrow\nonumber\\
     \hat{A}&=& \int dg
     \;f(g)\;\tilde{U}(g)\Leftrightarrow\nonumber\\
     \langle j^{\prime}m^{\prime}n^{\prime}|\hat{A}|jmn\rangle &=&
     N^{-1/2}_j\;\delta_{jj^{\prime}}\;
     \delta_{mm^{\prime}}\;f^j_{n^{\prime}n},\nonumber\\
     f^j_{n^{\prime}n}&=& N_j^{1/2}\int
     dg\;f(g)\;D^j_{n^{\prime}n}(g),\nonumber\\
     f(g)&=&\sum\limits_{jnn^{\prime}}\;N_j^{1/2}\;f^j_{n^{\prime}n}\;
     D^j_{nn^{\prime}}(g^{-1}) .
     \label{series}
     \end{eqnarray}

\noindent For such special operators $\hat{A}$ we in fact find:
     \begin{eqnarray}
     W_{\hat{A}}(g;jnn^{\prime}) &=&
     N_j^{-1/2}\;f^j_{n^{\prime}n},\nonumber\\
     \langle j^{\prime}m^{\prime}n^{\prime}|\hat{A}|jmn\rangle &=&
     \delta_{jj^{\prime}}\;\delta_{mm^{\prime}}\;W_{\hat{A}}(\cdot;
     jnn^{\prime}).
     \label{specialop}
     \end{eqnarray}

\noindent When the Weyl symbol of such an $\hat{A}$ is represented
as a block diagonal operator on ${\mathcal{H}}_0$ according to
(\ref{Ajtilde}), we have:
     \begin{eqnarray}
     \hat{A} &=& \int dg
     \;f(g)\;\tilde{U}(g)\Leftrightarrow\nonumber\\
     \tilde{A}(g)&=& g-\mbox{independent}=\int
     dg^{\prime}\;f(g^{\prime})\;{\mathcal {D}}_0(g^{\prime}).
     \label{Agind}
     \end{eqnarray}

\noindent Therefore when $\hat{A}$ on ${\mathcal{H}}$ is built up
exclusively from the operators of the right regular representation
$\tilde{U}(g)$, its Weyl symbol is the \underline{corresponding}
operator, in the sense of (\ref{Agind}), in the $SR$ of $G$,
stripping away the degeneracy of the regular representation. At
the generator level we can say that if $\hat{A}$ is a function
only of the generators $\tilde{J}_r$ of $\tilde{U}(g)$, then
$\tilde{A}$ is identically the same function of the generators of
the $SR$ ${\mathcal {D}}_0$ on ${\mathcal{H}}_0$. The block
diagonality of $\tilde{A}$ is of course assured.

This shows the important role of the $SR$ in the WWI for quantum
mechanics on a (compact simple) Lie group.

We return to the comment made after (\ref{tracetilde}) and ask
whether the definition of $\tilde{A}_j(g)$ for given $\hat{A}$
could have been altered so as to absorb the factors $N_j$
appearing on the right in that equation.  In that case that right
hand side would be expressible in terms of a trace over
${\mathcal{H}}_0$, which would make that relation more attractive.
However a careful analysis shows that in that case the simplicity
of the correspondence (\ref{Agind}) would be lost, and therewith
the direct relevance of the $SR$. Therefore to secure
(\ref{Agind}) we have to retain (\ref{tracetilde}) as it stands.
Ultimately this situation can be traced to the following source.
While the way in which the delta function in the trace relation
(\ref{traceortho}) appears is extremely elementary, when we
express it as in (\ref{Dproperties}) in terms of the irreducible
representation matrices of $G$ the dimensionality factors $N_j$
are essential.

\section{Concluding comments}

The method by which the $SR$ has been isolated within the regular
representation in the case of the group $SU(n)$ readily
generalises to all the other compact simple Lie group families,
namely $SO(2n)$, $SO(2n+1)$, $USp(2n)$ and even the five
exceptional groups. This is because in each case the concept of
highest weight in each $UIR$ is unambiguously defined, and
moreover the Lie algebra can be exhibited in the Cartan form, made
up of `shift' or`raising' and `lowering' generators in the
directions of the distinct root vectors. An interesting question
is how to effect a similar extraction of the $SR$ from the regular
representation in the case of finite groups, say the permutation
groups $S_N$. This presents interesting algebraic problems as
generators, shifts along root vectors etc. are no longer
available. The construction of the Schwinger representation for
the permutation groups $S_{n}$ has attracted attention in the
mathematical literature: see, for instance \cite{inglis}.

Two other general questions suggest themselves bearing in mind the
basic properties of the $SR$ : simple reducibility and
completeness: How are these properties reflected in the 'classical
limit', can one give some differential-geometric or
manifold-theoretic characterisations at the level of the coadjoint
orbit space of the Lie group? If one next takes the direct product
of the SR with itself, the simple reducibility aspect is likely to
change, yet one can ask if any simplifying features remain. We
hope to return to some of these questions elsewhere.


\begin{thebibliography}{99}
\bibitem{qamfoot} It is to be noted that in J.Schwinger, ref.\cite{schwinger},
eqn.(2.61), the ranges chosen are $0\leq \alpha, \gamma\leq
4\pi,\;0\leq \beta \leq \pi$, which amounts to covering $SU(2)$
twice.  In ref.\cite{qam}, the ranges chosen are $0\leq \alpha,
\gamma\leq 2\pi,\;0\leq\beta\leq \pi$ and $2\pi\leq \beta \leq
3\pi$.
\bibitem{miofoot} This leads to interesting consequences and structures
which are completely absent in the Cartesian case.
\end{thebibliography}
\end{document}